\RequirePackage{lineno}
\documentclass[twocolumn,letterpaper,aps,prc,longbibliography,superscriptaddress,nofootinbib,floatfix]{revtex4-2}

\usepackage{amsmath}	
\usepackage{graphicx}	
\usepackage{hyperref}
\hypersetup{colorlinks=true, urlcolor=cyan, linkcolor=blue, citecolor=blue}

\usepackage{xspace}	

\newcommand{\pt}{\mbox{$p_T$}\xspace}

\newcommand{\snn}{\mbox{$\sqrt{s_{_{NN}}}$}\xspace}
\newcommand{\sqsn}{\mbox{$\sqrt{s_{_{NN}}}$}\xspace}

\newcommand{\pp}{\mbox{$p$$+$$p$}\xspace}
\newcommand{\dau}{\mbox{$d$$+$Au}\xspace}

\newcommand{\pau}{\mbox{$p$$+$Au}\xspace}

\newcommand{\dphi}{\mbox{$\Delta\phi$}\xspace}

\newcommand{\heau}{\mbox{$^3$He$+$Au}\xspace}
\newcommand{\ppb}{\mbox{$p$$+$Pb}\xspace}

\newcommand{\pa}{\mbox{$p$$+$$A$}\xspace}

\newcommand{\pdheau}{\mbox{$p/d/^{3}$He$+$Au}\xspace}

\newcommand{\ttpc}{\mbox{3$\times$2PC}\xspace}

\begin{document}

\title{Checking Non-Flow Assumptions and Results via PHENIX Published Correlations in 
$p$$+$$p$, $p$$+$Au, $d$$+$Au, $^3$He$+$Au at $\sqrt{s_{NN}}$ = 200 GeV}

%

\newcommand{\colorado}{University of Colorado, Boulder, Colorado 80309, USA}
\affiliation{\colorado}

\newcommand{\uncg}{University of North Carolina, Greensboro, North Carolina 27413, USA}
\affiliation{\uncg}

\newcommand{\pusan}{Pusan National University, Busan, 46241, South Korea}
\affiliation{\pusan}

\author{J.L. Nagle} \affiliation{\colorado}
\author{R. Belmont} \affiliation{\uncg}
\author{S.H. Lim} \affiliation{\pusan}
\author{B. Seidlitz} \affiliation{\colorado}

\date{\today}

\begin{abstract}

Recently the PHENIX Collaboration has made available two-particle correlation Fourier coefficients for multiple detector combinations in minimum bias \pp and 0--5\% central \pau, \dau, and \heau collisions at \snn = 200 GeV~\cite{new:phenix}.    Using these coefficients for three sets of two-particle correlations, azimuthal anisotropy coefficients $v_2$ and $v_3$ are extracted for midrapidity charged hadrons as a function of transverse momentum.    
In this paper, we use the available coefficients to explore various non-flow hypotheses as well as compare the results with theoretical model calculations.  
The non-flow methods fail basic closure tests with \textsc{ampt} and \textsc{pythia/angantyr}, particularly when including correlations with particles in the low multiplicity light-projectile going direction.    In data, the non-flow adjusted $v_2$ results are modestly lower in \pau and the adjusted $v_3$ results are more significantly higher in \pau and \dau.   However, the resulting higher values for the ratio $v_3/v_2$ in \pau at RHIC compared to \ppb at the LHC is additional evidence for a significant over-correction.
Incorporating these additional checks, the conclusion that these flow coefficients are dominated by initial geometry coupled with final-state interactions (e.g.~hydrodynamic expansion of quark-gluon plasma) remains true, and explanations based on initial-state glasma are ruled out.     The detailed balance between intrinsic and fluctuation-driven geometry and the exact role of weakly versus strongly-coupled pre-hydrodynamic evolution remains an open question for triangular flow, requiring further theoretical and experimental investigation.   

\end{abstract}


\maketitle

\section{Introduction}
 The standard time evolution model for heavy ion collisions involves multiple stages, similar to the stages of the standard big bang time evolution model for the universe.   In collisions between large nuclei, for example Au+Au at the Relativistic Heavy Ion Collider (RHIC) and Pb+Pb at the Large Hadron Collider (LHC), the epochs proceed with (i) the initial collision including hard scattering of partons, (ii) a pre-hydrodynamic phase, (iii) an extended hydrodynamic phase where the matter has a temperature exceeding $T>$~155~MeV and is thus considered a quark-gluon plasma (QGP), and finally after hadronization (iv) a stage of hadronic scattering until the densities are low enough that the particles are free streaming~\cite{Heinz:2013th}.   For more than a decade, the success of hydrodynamics was assumed to imply near-equilibration/thermalization and hence epoch (ii) was referred to as pre-equilibrium.    It is now generally recognized that the collision system with rapid longitudinal expansion does not achieve equilibrium~\cite{Romatschke:2016hle}, and this stage has been re-named pre-hydrodynamization, viz.~the time before hydrodynamics applies---for further discussion see Refs.~\cite{Heller:2015dha,Romatschke:2017ejr}.   Constraining properties of the QGP, such as shear and bulk viscosity and the nuclear equation-of-state, requires a modeling of all these stages combined with detailed, quantitative comparison with experimental data---see Ref.~\cite{Nijs:2020roc,Everett:2020xug} for recent examples.
 
 Similar experimental signatures for QGP formation and the standard time evolution model have also been observed in smaller collision systems such as \pau, \dau, \heau at RHIC and \pp, \ppb at the LHC---see
 Ref.~\cite{Nagle:2018nvi} for a recent review.    However, in contrast to the larger QGP droplets formed in A+A collisions, if smaller QGP droplets are indeed formed in smaller collision systems, the time spent in that epoch is significantly shorter, thereby enabling the pre-hydrodynamic physics to play a larger role.    
 In Ref.~\cite{Romatschke:2015gxa}, the author specifically details how light-heavy ion collisions can provide a window into pre-hydrodynamic QCD evolution.    However, these short lived systems are also more sensitive to the first stage (i) in terms of both the initial geometry and local (typically few-particle) correlations between particles, the latter of which is often referred to as non-flow.    

A specific proposal was put forth to collide proton, deuteron, and helium-3 projectiles on nuclear targets at RHIC, utilizing the unique capabilities of that facility, to discern whether ``flow-like'' patterns are indeed attributable to mini-QGP droplet formation~\cite{Nagle:2013lja}.   The PHENIX Collaboration has published a series of papers culminating in the Nature Physics paper with elliptic ($v_2$) and triangular ($v_3$) coefficients as a function of transverse momentum (\pt) in \pau, \dau, and \heau collisions~\cite{PHENIX:2018lia}.  Theoretical predictions within the evolution model including a short-lived QGP droplet describe the data quantitatively.   In contrast, a long-standing question of whether initial-state correlations from stage (i) via exotic glasma diagrams could describe the data has been settled.   These glasma correlations are not able to describe the magnitudes of $v_2$ or $v_3$, nor their \pt dependence, nor the projectile nuclear size dependence~\cite{Mace:2018vwq,Mace:2018yvl}.  However, the influence of other epoch (i) non-flow correlations remains a topic of active discussion.


The PHENIX Collaboration has recently published new results on $v_2$ and $v_3$ in \pau,
\dau, and \heau collisions---in excellent agreement with
Ref.~\cite{PHENIX:2018lia}---using three sets of two-particle correlations (called the
\ttpc method)~\cite{new:phenix}.  In the \ttpc method, determination of the $v_n$ requires
three sets of two-particle correlations and the extraction of the Fourier coefficients
$c_n$ of each set.  The publication of the $c_{n}$ coefficients wth statistical
uncertainties for all two-particle correlation combinations in Ref.~\cite{new:phenix}
provides ample opportunity for additional exploration of the data and interpretations
thereof.

\section{Non-Flow Subtraction Methods}

We utilize three methods to estimate and subtract non-flow contributions in a given correlation function of interest.    These methods assume that the shape the non-flow contribution to the correlation function is multiplicity and collision-system independent. In this study, we use the correlation functions from \pp collisions.     The first method, called the $c_{1}$-method, estimates the non-flow contributions to $c_{n}$ coefficients of the correlation function of interest by scaling the $c_{n}$ coefficients in \pp collisions ($c_{n}^{pp}$) by the ratio of $c_{1}$ coefficients,

\begin{align}
    c_{n}^\mathrm{corrected} = c_{n} - c_{n}^{pp}\times \frac{c_{1}}{c_{1}^{pp}}.
\end{align}
This method assumes the $c_1$ coefficient is purely from non-flow effects, and that there is no flow contribution to the higher coefficients in  the low-multiplicity  reference, in this case \pp collisions.


We also use a template fit method developed by the ATLAS collaboration~\cite{Aad:2019igg}.
In the template fit method, a correlation function from high multiplicity $C(\dphi)$ is
described by a scaled correlation function from \pp collisions and an additional flow
contribution,
\begin{align}
        C(\dphi) &= FC^{pp}(\dphi) + C^\mathrm{flow}(\dphi)\nonumber\\ &=
        FC^{pp}(\dphi) + G \left( 1 + \sum^{\infty}_{n=2} 2 c_{n} \cos(n\dphi) \right),
\end{align}
where $F$ and $c_{n}$ are determined by the fitting procedure, and $G$ is fixed by
requiring the integrals of $C(\dphi)$ and $C^\mathrm{flow}(\dphi)$ to be equal.

Finally we use the zero-yield-at-minimum (ZYAM) method~\cite{Adare:2008ae}, where one assumes that the number of correlated pairs is zero at the correlation function minimum.  This minimum is then subtracted out to obtain the distribution of the correlated per-trigger-yields.   Application of these correlated per-trigger-yields for the non-flow adjustment is detailed in Ref.~\cite{Aad:2014lta}.  In all methods, the extracted $c_{n}$ coefficients are a multiplication of flow coefficients of two particles, $c_{n}=v_{n,a}\times v_{n,b}$ which allow for the measurement of the single particle $v_n$ coefficients with three sets of correlations.     

Comparing results between methods is instructive; however, it should be kept in mind that all have a subset of assumptions that are common and thus the range of results is not automatically a good proxy for a systematic uncertainty.  More detailed discussion on the non-flow subtraction methods can be found in Ref.~\cite{Lim:2019cys}.    

\clearpage

\section{Data Results}

\begin{figure*}[!ht]
    \centering
    \includegraphics[width=0.9\linewidth]{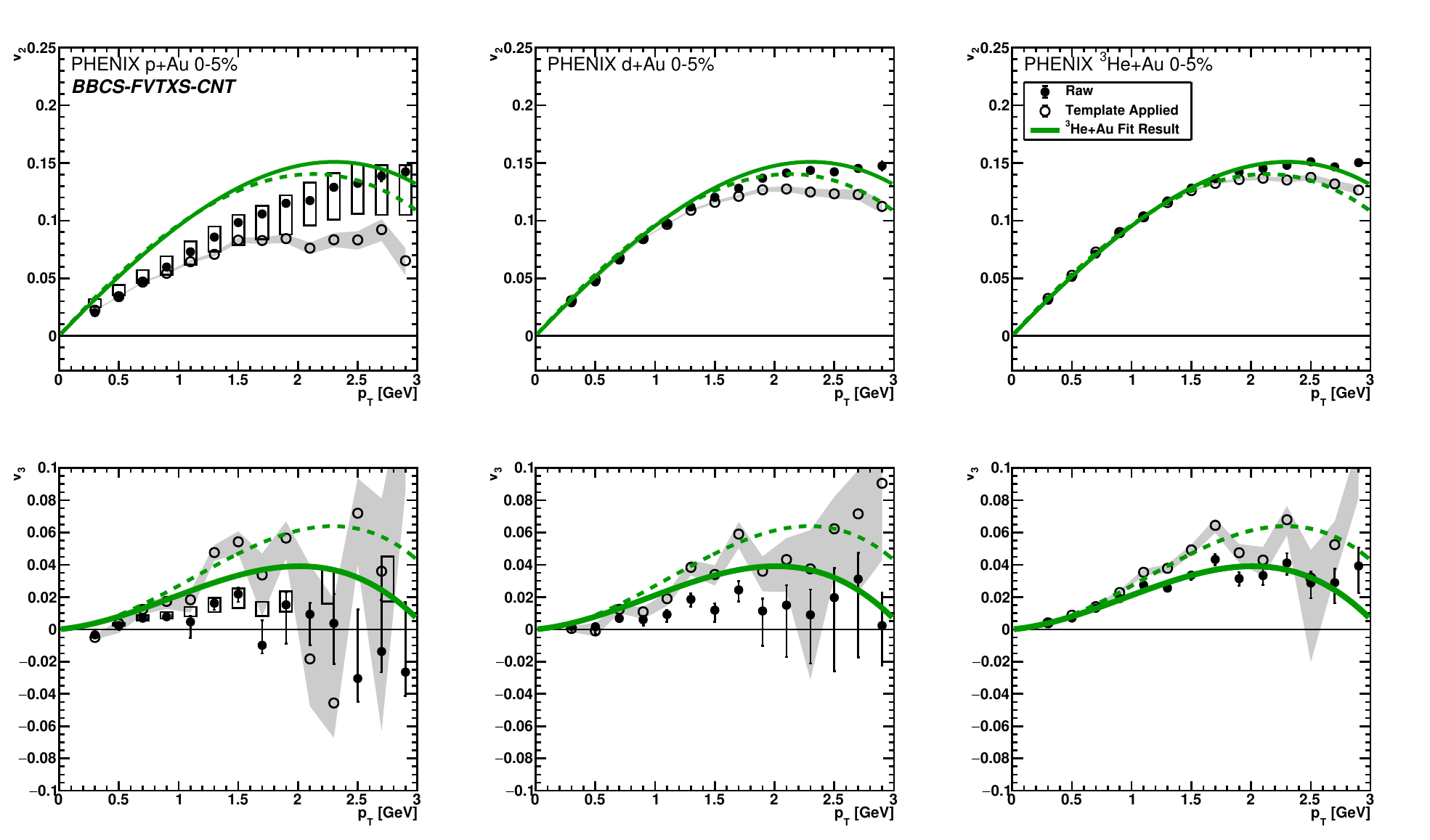}
    \caption{PHENIX published $v_2$ (top) and $v_3$ (bottom) for \pau, \dau, and \heau 0--5\% collisions at \sqsn = 200 GeV from left to right, respectively.  Solid points are the raw values and the open points are template-adjusted.   Non-flow systematic uncertainties estimated by the PHENIX Collaboration in Ref.~\cite{PHENIX:2018lia} are shown for \pau $v_2$ and $v_3$ results as examples.   The solid and dashed lines are fits to the \heau results.}
    \label{fig:phenix_template}
\end{figure*}
\begin{figure*}[!ht]
    \centering
    \includegraphics[width=0.9\linewidth]{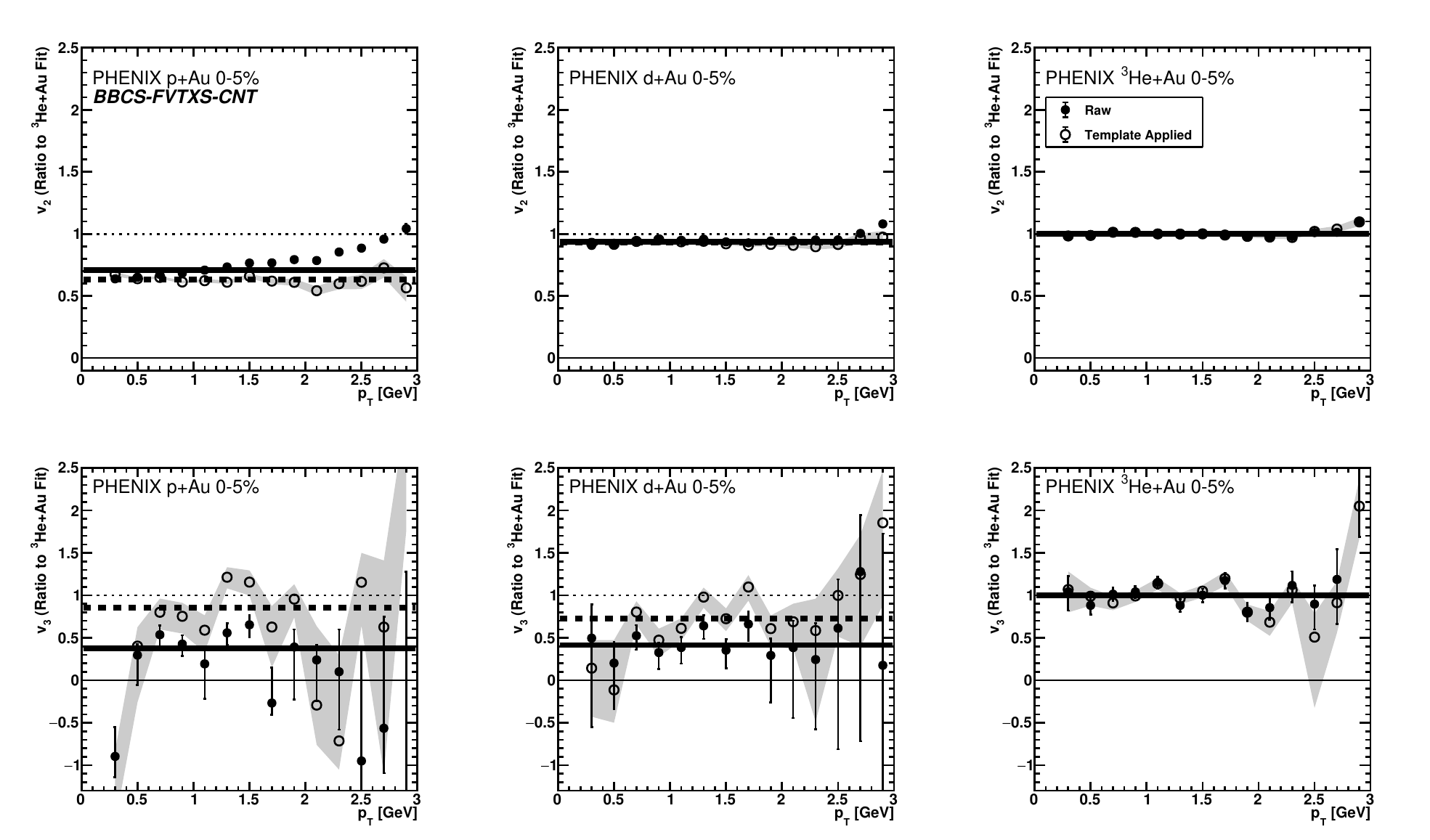}
    \caption{PHENIX published $v_2$ (top) and $v_3$ (bottom) for \pau, \dau, and \heau 0--5\% collisions at \sqsn = 200 GeV from left to right, respectively, as a ratio to the \heau fit functions.   Solid points are the raw values and the open points are template-adjusted.   Simple horizontal level fits are shown as solid and dashed lines for the two cases, respectively.}
    \label{fig:phenix_template_ratio}
\end{figure*}

\begin{figure*}[!ht]
    \centering
    \includegraphics[width=0.9\linewidth]{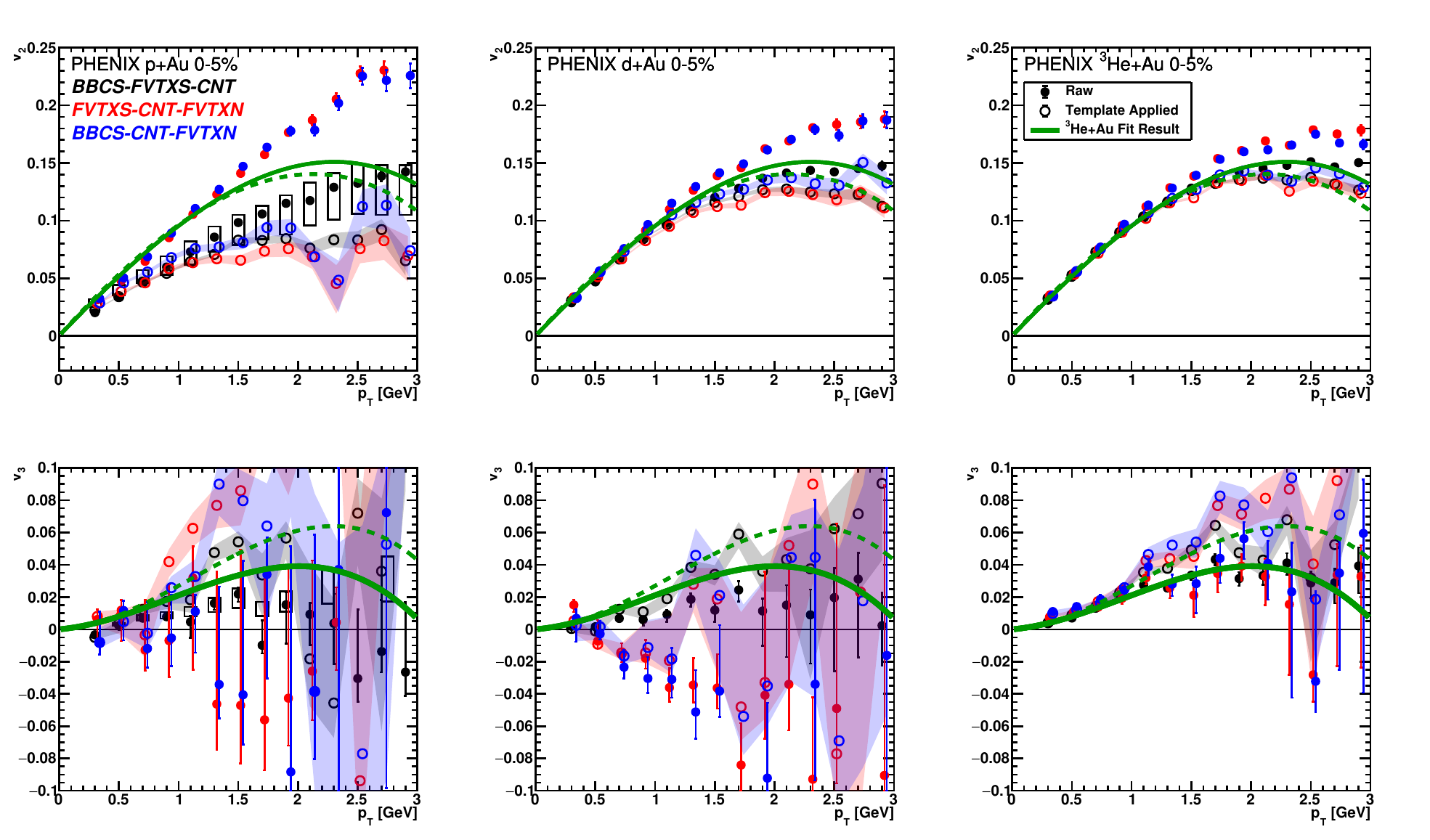}
    \caption{PHENIX published $v_2$ (top) and $v_3$ (bottom) for \pau, \dau, and \heau 0--5\% collisions at \sqsn = 200 GeV from left to right, respectively, from three different sets of detector combinations.   Solid points are the raw values and the open points are template-adjusted.   Non-flow systematic uncertainties estimated by the PHENIX Collaboration in Ref.~\cite{PHENIX:2018lia} are shown for \pau $v_2$ and $v_3$ results as an example.   The solid and dashed lines are fits to the \heau results.}
    \label{fig:phenix_template99}
\end{figure*}
\begin{figure*}[h]
    \centering
    \includegraphics[width=0.9\linewidth]{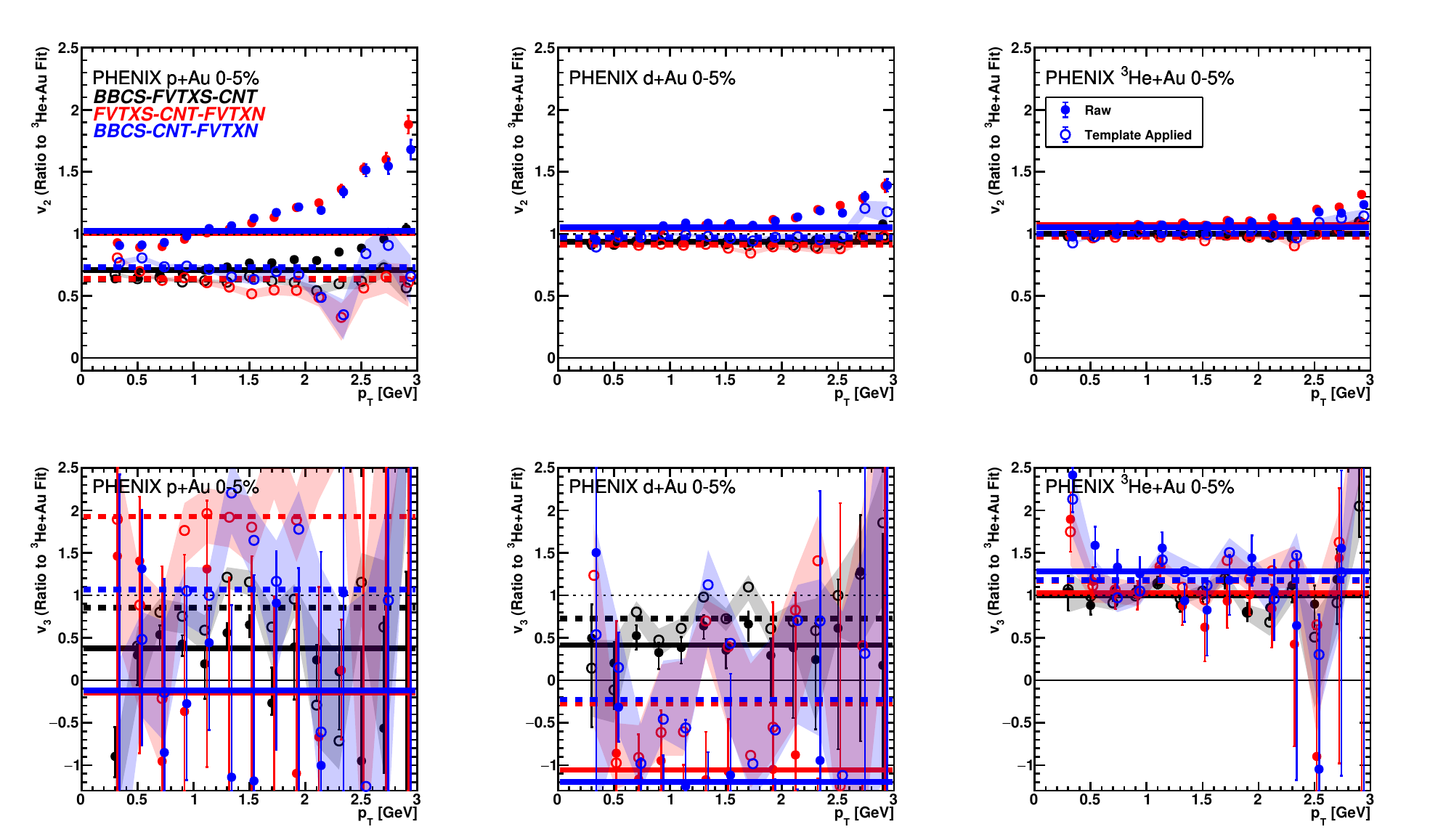}
    \caption{PHENIX published $v_2$ (top) and $v_3$ (bottom) for \pau, \dau, and \heau 0--5\% collisions at \sqsn = 200 GeV from left to right, respectively, as a ratio to the \heau fit functions, from three difference sets of detector combinations.   Solid points are the raw values and the open points are template-adjusted.
    Simple horizontal level fits are shown as solid and dashed lines for the two cases, respectively.}
    \label{fig:phenix_template99_ratio}
\end{figure*}

\begin{figure*}[h]
    \centering
    \includegraphics[width=0.9\linewidth]{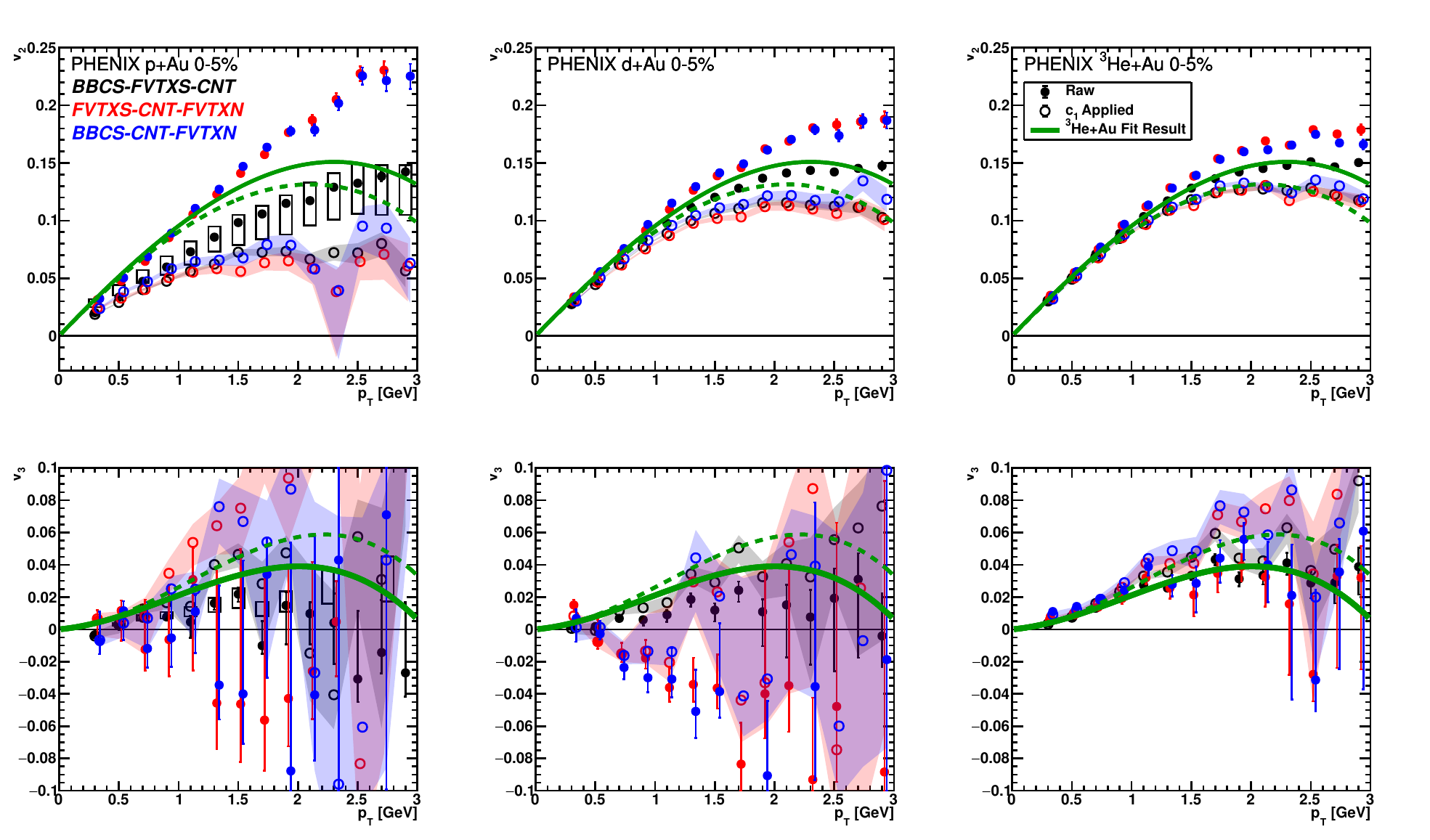}
    \caption{PHENIX published $v_2$ (top) and $v_3$ (bottom) for \pau, \dau, and \heau 0--5\% collisions at \sqsn = 200 GeV from left to right, respectively.   Solid points are the raw values and the open points are $c_{1}$-method-adjusted.   Non-flow systematic uncertainties estimated by the PHENIX Collaboration in Ref.~\cite{PHENIX:2018lia} are shown for \pau $v_2$ and $v_3$ results as examples.   The solid and dashed lines are fits to the \heau results.}
    \label{fig:phenix_c1}
\end{figure*}
\begin{figure*}[h]
    \centering
    \includegraphics[width=0.9\linewidth]{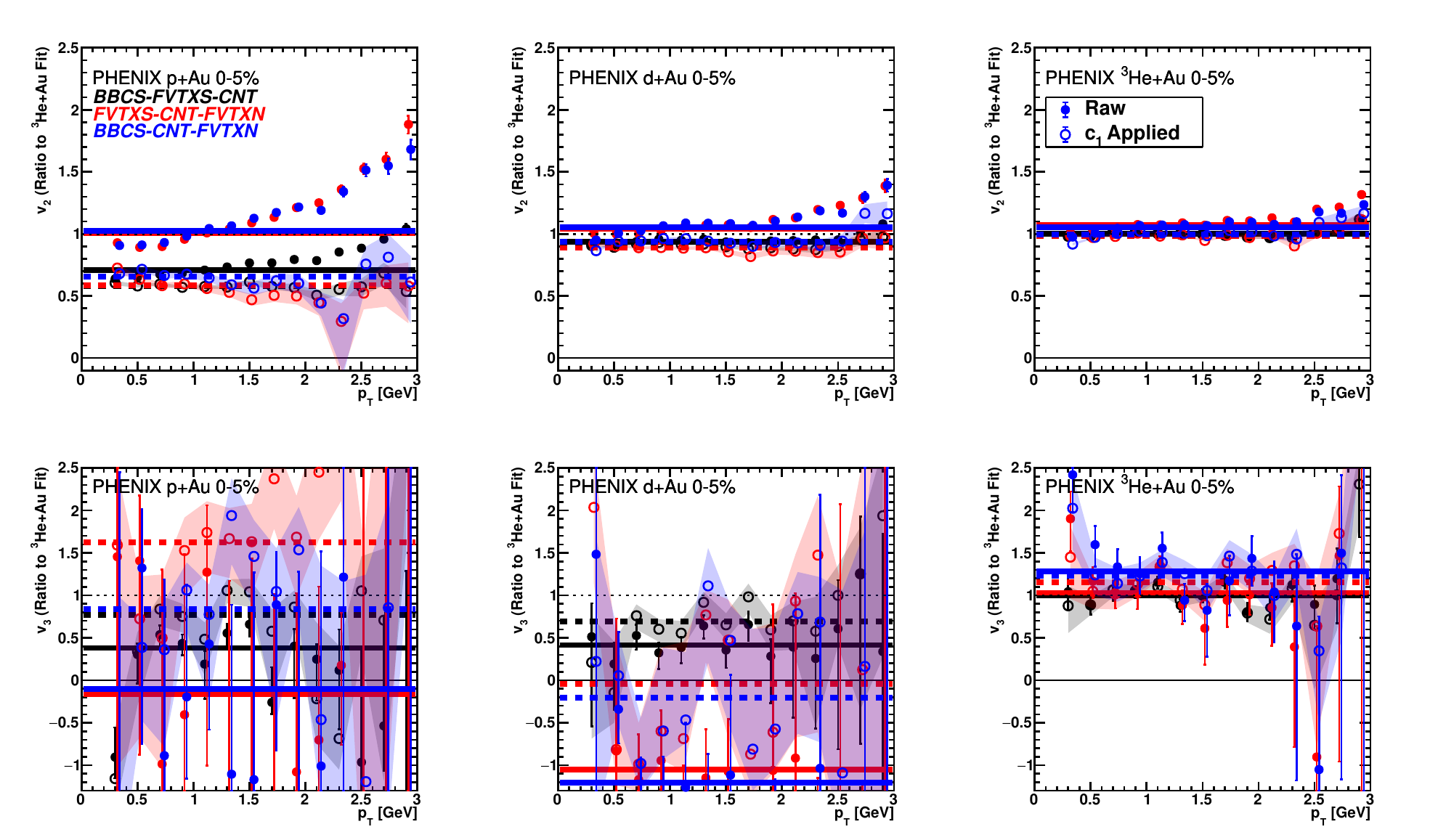}
    \caption{PHENIX published $v_2$ (top) and $v_3$ (bottom) for \pau, \dau, and \heau 0--5\% collisions at \sqsn = 200 GeV from left to right, respectively, as a ratio to the \heau fit functions.   Solid points are the raw values and the open points are $c_{1}$-method-adjusted.}
    \label{fig:phenix_c1_ratio}
\end{figure*}

   The PHENIX detectors utilized in this analysis are the PHENIX Beam-Beam Counter South (BBCS) covering $-3.9 < \eta < -3.1$, the Forward Vertex Tracker South (FVTXS) covering $-2.2 < \eta < -1.2$, the Central Tracker (CNT) covering $-0.35 < \eta < 0.35$, the Forward Vertex Tracker North (FVTXN) covering $1.2 < \eta < 2.2$, and the Beam-Beam Counter North (BBCN) covering $3.1 < \eta < 3.9$.   
   
   Figure~\ref{fig:phenix_template} shows as closed points the PHENIX experimental results for $v_2$ (top) and $v_3$ (bottom) as a function of \pt at midrapidity for \pau (left), \dau (middle) and \heau (right) central 0--5\% collisions at \sqsn = 200 GeV.   These results are extracted via three sets of two-particle correlations---between the BBCS-FVTXS, BBCS-CNT, and FVTXS-CNT---and thus labeled as BBCS-FVTXS-CNT.    This combination was chosen since it has the largest rapidity gaps between detectors and the BBCS and FVTXS, being in the Au-going direction, have the highest multiplicities thus minimizing non-flow contributions.    The uncertainties for the \ttpc results are statistical only.
   The PHENIX experiment estimated non-flow contributions in the earlier event plane analysis and included them as asymmetric systematic uncertainties~\cite{PHENIX:2018lia}.    However this systematic uncertainty was only determined in the event plane method from the BBCS-CNT correlation---see Ref.~\cite{PHENIX:2018lia} for details.   We overlay those systematic uncertainties for the \pau $v_2$ and $v_3$ example for visual comparison purposes, noting that the \pt binning is slightly different from the \ttpc results.   It is only with the new analysis via \ttpc~\cite{new:phenix} that one can apply the non-flow method to all contributions as detailed below. 

Figure~\ref{fig:phenix_template} also shows as open points the adjusted results when applying the template fit method, as detailed above.   The low multiplicity reference comes from the PHENIX \pp minimum bias data, which corresponds to 55$\pm$5\% of the inelastic cross section of 42 mb~\cite{Adare:2013nff}.     
The template fit method is applied individually to each of the three sets of two-particle correlations.   

As the $c_{2}$ coefficients are positive in \pp, the adjusted \pdheau $c_{2}$ coefficients decrease.   However, in calculating the $v_2$, two correlation coefficients are in the numerator and one in the denominator, and hence the direction of the adjustment to $v_2$ is non-trivial.    In contrast, the $c_{3}$ coefficients are negative in \pp, and hence the adjusted \pdheau $c_{3}$ coefficients increase.   For the same reason as for $v_2$, the direction of the adjustment to $v_3$ is non-trivial.  In the end, however, the $v_2$ ($v_3$) values decrease (increase) in all cases, in line with expectations.    The grey bands indicate the statistical uncertainties only on the template-adjusted values resulting from the statistical uncertainties on the coefficients in \pdheau and the \pp reference.   There are some cases where the combination of $c_{n}$ coefficients results in an imaginary $v_{n}$, and these are plotted as negative values on the vertical scale.
The solid (dashed) lines are fits to the \heau raw (template-adjusted) results and are shown in all panels for comparison.   The template-adjusted \pau results are consistent with the raw results within the original non-flow estimated uncertainty for \pt $< 2$~GeV. 
The template-adjusted results for the $v_3$ are higher than the raw $v_3$ beyond the original non-flow estimated uncertainties.

Figure~\ref{fig:phenix_template_ratio} shows the ratio of the points from Figure~\ref{fig:phenix_template} relative to the \heau fits, and hence by construction the right most panel values are consistent with one.   One observes a slightly lower $v_2$ in \dau relative to \heau, of order 5--10\%, and independent of \pt regardless of the template-adjustment.     The $v_2$ in \pau relative to \heau is slightly lower with the template-adjustment and appears to remove the \pt dependence above 2~GeV seen in the raw results.   For the $v_{3}$, the \pau and \dau raw results are consistently 60\% lower than the \heau and flat with \pt within uncertainties.   The template-adjusted results are higher and more consistent with only 15--30\% lower than the \heau results.    As before, the grey bands are propagated statistical uncertainties only.  

Figures~\ref{fig:phenix_template99} and \ref{fig:phenix_template99_ratio} show the $v_{2,3}$ and ratios to the \heau fits but now for two additional sets of detector combinations (FVTXS-CNT-FVTXN and BBCS-CNT-FVTXN).    The non-flow contributions are expected to be larger in the FVTXN due to the significantly lower multiplicity in the light-projectile-going direction and the lower expected flow---see Ref.~\cite{Adare:2018toe} for example.  The raw $v_2$ results from both new combinations are higher than the BBCS-FVTXS-CNT combination discussed previously, most strikingly so in the \pau case.    The template-adjusted results for different detector combinations are in better agreement with each other, though a 15--25\% relative difference remains in the \pau case.    

In the $v_3$ case, the combinations involving the FVTXN result in imaginary raw $v_3$ values in \pau and \dau, plotted as negative values for visualization.   The template-adjusted values shift significantly upward into the positive, real range for many data points; however, the three detector combinations of values remain disparate in the \pau and \dau cases.

We also show the three detector combinations and the $c_1$-method adjustment in Figures~\ref{fig:phenix_c1} and ~\ref{fig:phenix_c1_ratio}.    The results are qualitatively similar to the template method. However, the adjustment is somewhat larger, as can be seen in the \pau $v_2$ values, which are now slightly below the systematic uncertainties of the PHENIX data even for $p_T<$~2~GeV.    Since the $c_1$-method assumes no flow in the \pp reference, this larger non-flow adjustment is in line with expectations.    Results with the ZYAM method (not shown) are qualitatively similar to the template and $c_1$-method adjustments.


\clearpage

\section{Non-Flow Checks}

One method to check the non-flow adjustment methods is with Monte Carlo calculations.   Applications of the methods to Monte Carlo calculations without final-state interactions, i.e., flow, have been done with \textsc{pythia}, \textsc{pythia/angantyr}, \textsc{hijing} and on \textsc{ampt} which also includes final-state partonic and hadronic scattering---see Ref.~\cite{Lim:2019cys} for example.    Here we utilize two of these calculations to test the non-flow adjustment methods utilizing the PHENIX kinematic selections.

\begin{figure}[!ht]
    \centering
    \includegraphics[width=0.85\linewidth]{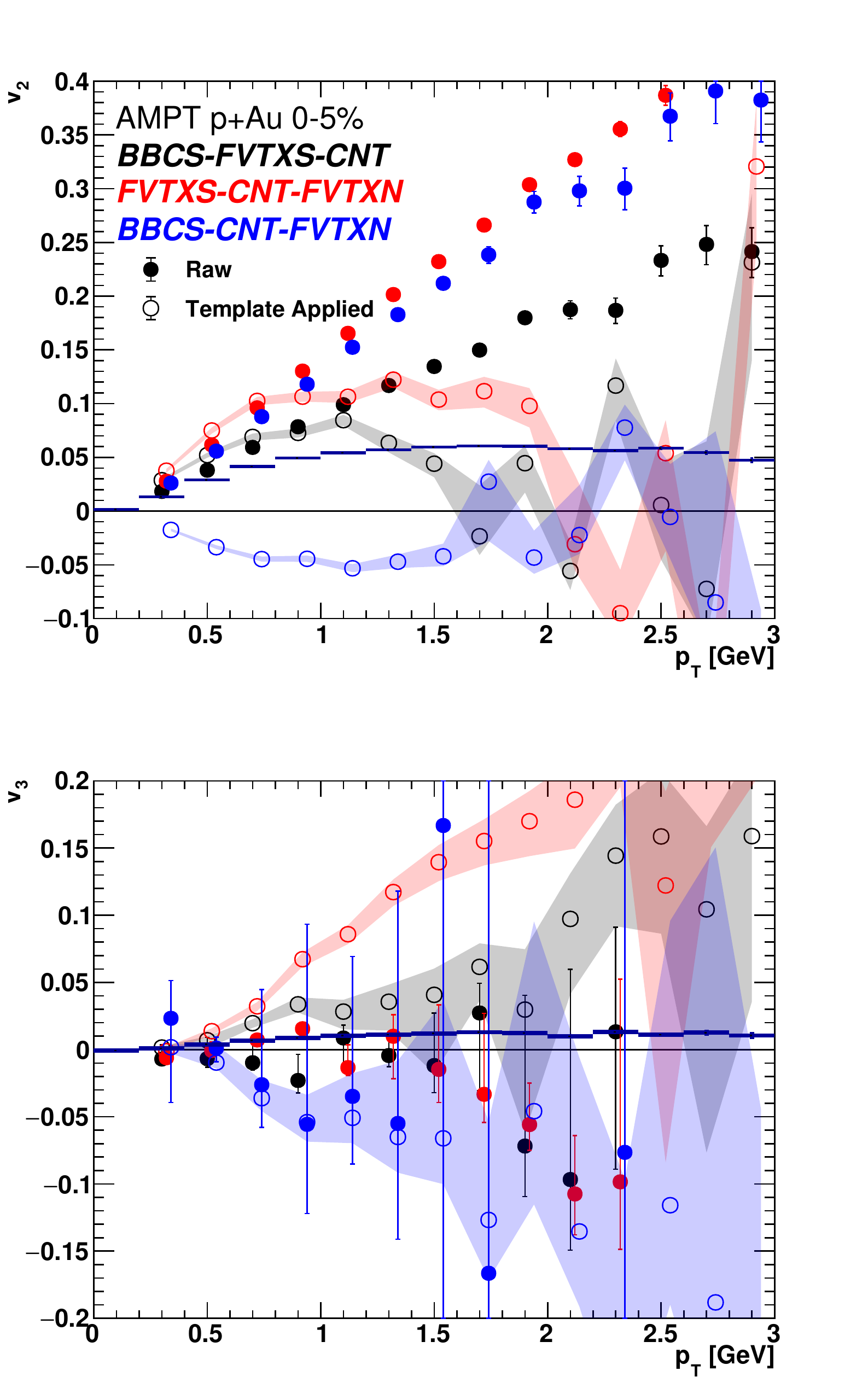}
    \caption{\textsc{ampt} $v_2$ (top) and $v_3$ (bottom) for \pau central ($b<2$~fm) collisions at \sqsn = 200 GeV from from three different sets of detector combinations.   Solid points are the raw values and the open points are the template-adjusted values.   The solid horizontal lines correspond to the ``truth'' result as detailed in the text.   Negative values correspond to imaginary results when one or more of the coefficients in the square-root are negative.}
    \label{fig:ampt_c1}
\end{figure}

A Multi-Phase Transport (\textsc{ampt}) is a non-hydrodynamic framework for calculating the various epochs in the heavy ion time evolution~\cite{Lin:2004en}.   For the initial stage (i) the geometry and initial color strings come from the \textsc{hijing} model, the next stage (ii) is essentially free streaming modeled by a formation time for the partons, (iii) is modeled via on-shell parton scattering in the \textsc{zpc} framework, and (iv) is modeled with the \textsc{arc} hadron scattering package.  We have run 100 million minimum bias \pp and central (impact parameter $b < 2$~fm) \pau  events and analyzed the Monte Carlo data via the same sets of three detector combinations (in terms of \pt and pseudorapidity coverage) as used in the PHENIX analysis.    The results of using the correlation coefficients to calculate the raw $v_{n}$ and the template-method adjusted $v_{n}$ are shown in Figure~\ref{fig:ampt_c1}.

Similar to the PHENIX data, the raw \textsc{ampt} $v_2$ values are significantly higher with the FVTXS-CNT-FVTXN and BBCS-CNT-FVTXN combinations, presumably due to larger non-flow contributions; however, all the $v_2$ values are almost a factor of two higher than the PHENIX data.    In the case of $v_3$, the BBCS-FVTXS-CNT result is near zero, while the other combinations yield imaginary results (shown as negative values).    Also shown are the template-adjusted values which have significant variations between detector combinations for both $v_2$ and $v_3$.    The large imaginary result 
in the BBCS-CNT-FVTXN combination results from the $c_n$ sign in the BBCS-FVTXN combination.
The solid dashed lines are the ``truth'' result calculated as the hadron $v_{n}$ relative to the $\Psi_{n}$ defined from the initial parton geometry---see Ref.~\cite{Koop:2015wea} for details on the method.   We highlight that this method may not correspond to the absolute ``truth'' expectation even if non-flow is perfectly accounted for.   However, it is clear that the \textsc{ampt} results do not show good closure, i.e., a validation of the methodology.    The $c_{1}$ and ZYAM methods yield qualitatively similar conclusions.  

The failure of the non-flow adjustment methods is not unexpected in the \textsc{ampt} case.    First, from tracing the parton scattering history, the hard scattered partons do scatter further with medium partons.   This means that the ``jet shape'' explicitly changes between \pp and \pau collisions, and thus violates one of the basic assumptions in all of the non-flow methods~\cite{Nagle:2018eea}.    Additionally, the modeling of the initial stage by \textsc{hijing} results in a near-side jet correlation that has been observed to be wider than in real \pp data.    This additional non-flow contribution has been demonstrated to lead to adjustment failures~\cite{Lim:2019cys}.   

\begin{figure}[!ht]
\centering
\includegraphics[width=0.85\linewidth]{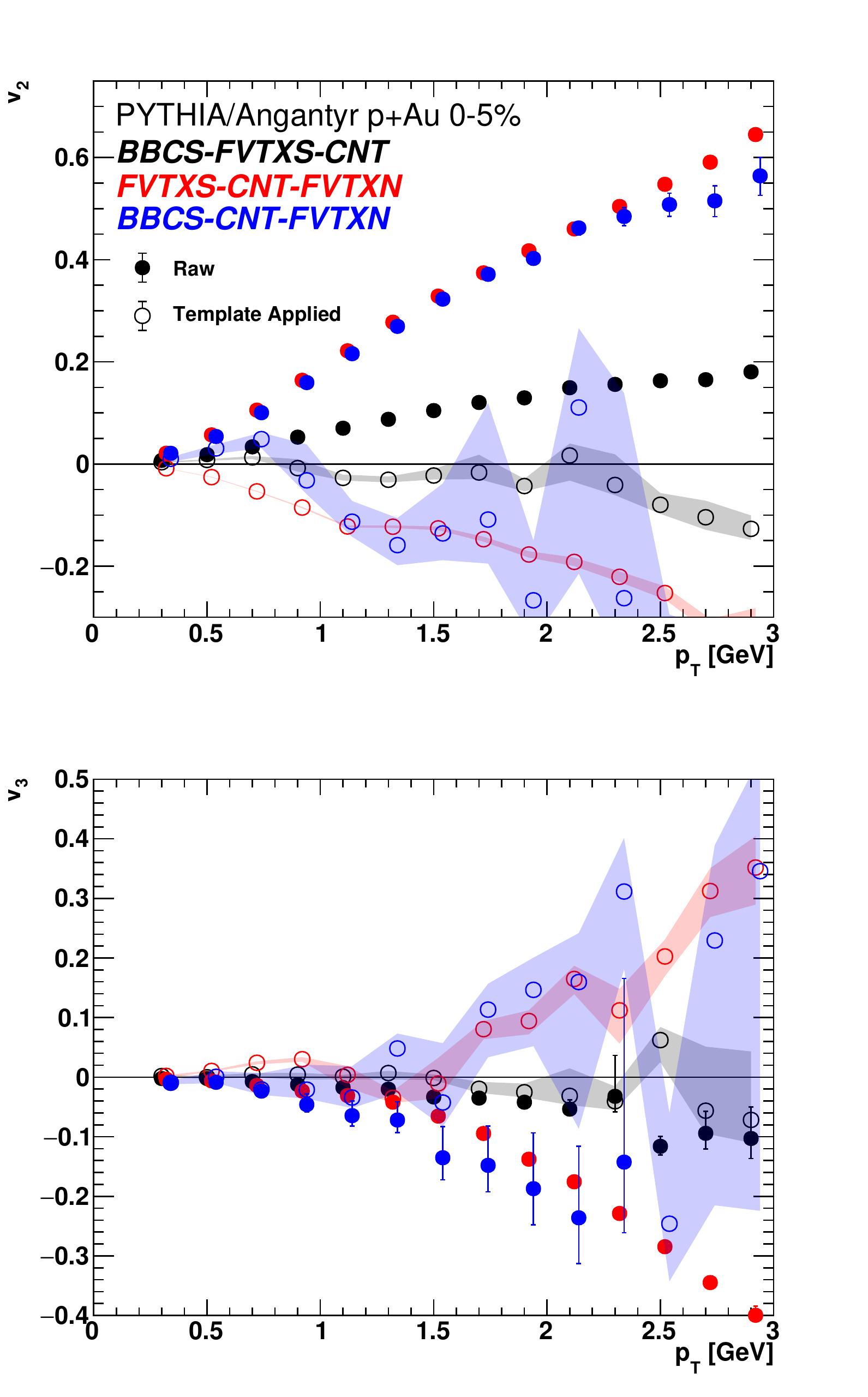}
\caption{\textsc{pythia/angantyr} $v_2$ (top) and $v_3$ (bottom) for \pau central ($b<2$~fm) collisions at \sqsn = 200 GeV from from three different sets of detector combinations.   Solid points are the raw values and the open points the template-adjusted values.   Negative values correspond to imaginary results when one or more of the coefficients in the square-root are negative.}
\label{fig:pythia_c1}
\end{figure}

Next we check the methods with the \textsc{pythia/angantyr} Monte Carlo~\cite{Bierlich:2018xfw}.   The results are shown in Figure~\ref{fig:pythia_c1}, and indicate a large non-flow contribution with the largest for detector combinations including the FVTXN.    The $v_3$ coefficients are imaginary in all cases.   The non-flow adjusted results show a significant over-correction in the case of the FVTXN detector combinations.   Results from the combination with the smallest non-flow (BBCS-FVTXS-CNT) have adjusted results reasonably close to zero, though still with a residual non-closure (i.e. a non-zero extraction of final-state $v_n$). The $c_{1}$ and ZYAM methods yield qualitatively similar conclusions.  

\section{Discussion}

The above calculations are mathematically well-defined.   The question is whether the raw results or the various non-flow adjusted results are reliable in a way that experiments can define ``experimental quantities'' with well-constrained uncertainties.  One clear takeaway message is that the assumptions of the non-flow methods are always violated, and the question is how much are they violated and how big is the correction relative to these problems~\cite{Lim:2019cys}.

In the case of $v_{2}$ for the \pau, \dau, and \heau systems, the template, $c_1$-method, and  ZYAM adjustments yield reasonable (10--25\% level) agreement between detector combinations.   The \pau results are reasonably consistent with the raw results with asymmetric non-flow systematic uncertainties up to \pt = 2~GeV.   The non-flow adjustment is particularly prone to over-correction at higher \pt as demonstrated above.

However, the $v_3$ results in \pau and \dau are almost a factor of two higher relative to \heau in the template, $c_1$-method, and ZYAM adjusted cases compared with the raw case for the BBCS-FVTXS-CNT detector combination.    These differences are larger than the PHENIX published non-flow asymmetric systematic uncertainty~\cite{PHENIX:2018lia}.   For the other two detector combinations, the $v_3$ is imaginary for the raw values and receives a very large adjustment from the non-flow methods and with very large statistical uncertainties.

\begin{figure*}[!h]
\centering
\includegraphics[width=0.99\linewidth]{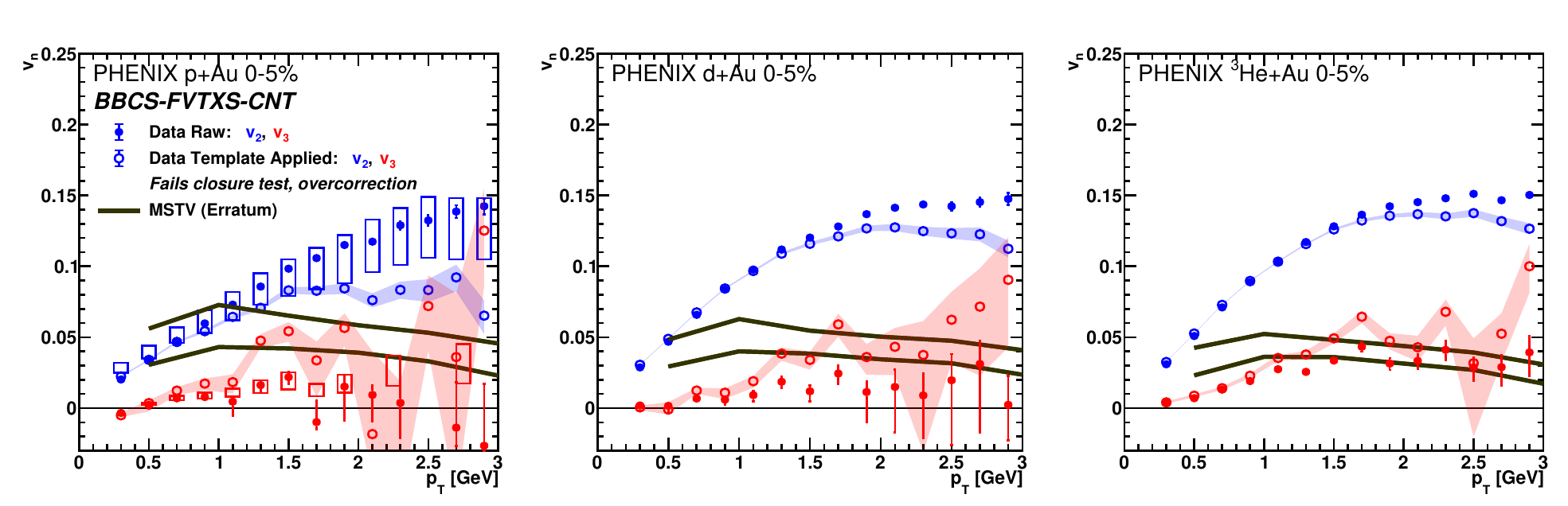}
\hfill
\includegraphics[width=0.99\linewidth]{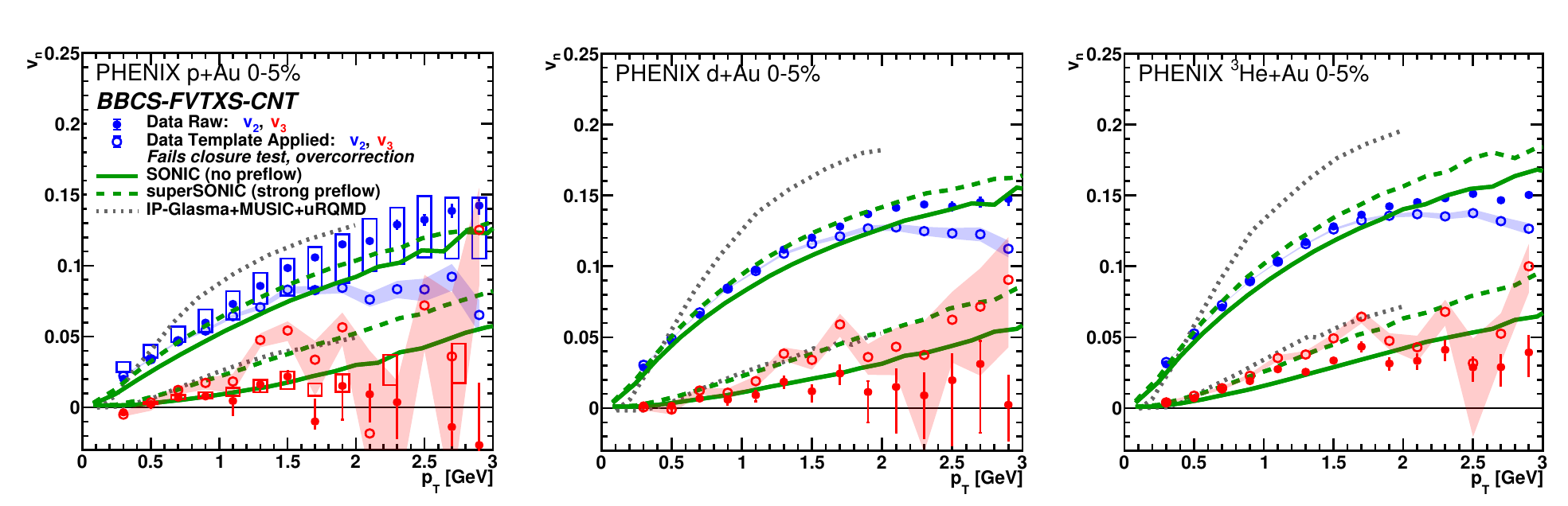}
\caption{PHENIX raw and template correction applied data $v_{2}$ and $v_{3}$  as a function of \pt in central \pau, \dau, and \heau collisions.   Also shown are theoretical calculations (initial-state glasma results in the upper panels and final-state hydrodynamic results in the lower panels) detailed in the text.}
\label{fig:uber}
\end{figure*}

Focusing on the case with the smallest non-flow contributions, Figure~\ref{fig:uber} shows the PHENIX raw data and template-adjusted data for $v_{2}$ and $v_{3}$ in the three collisions systems, \pau, \dau, and \heau for the BBCS-FVTXS-CNT detector combination.   In the upper panel, overlaid are calculations from the authors MSTV in the initial-state glasma framework~\cite{Mace:2018vwq,Mace:2018yvl}.     The calculations fail to describe the data with or without the non-flow adjustment. 
In the lower panel, overlaid are hydrodynamic calculations with the \textsc{sonic} (no pre-flow) and \textsc{supersonic} (with strong pre-flow) models~\cite{Romatschke:2015gxa} and the  \textsc{ip-glasma+music+urqmd} model~\cite{Schenke:2020mbo}.
As an aside, there are significantly larger theoretical uncertainties in the \textsc{ip-glasma+music+urqmd}~\cite{Schenke:2020mbo} case due to various choices in matching conditions between the weakly coupled pre-hydrodynamic (\textsc{ip-glamsa}) and hydrodynamic (\textsc{music}) stages, which does not exist in \textsc{supersonic} where both stages are strongly coupled.
With or without non-flow adjustment, the $v_2$ ordering between \pdheau systems follows expectations from initial geometry differences and final-state interactions.    The $v_3$ is lower in \pau and \dau compared with \heau, but the degree depends highly on the treatment of non-flow.
Hence the conclusions regarding the general agreement with geometry ordering and hydrodynamic modeling and the ruling out of initial-stage glasma correlations are retained.   

Despite the general agreement with hydrodynamic calculations, there are important physics implications for the variation in $v_{3}$ in \pau and \dau collisions. One important open issue in the field is the nature of the initial geometry in small systems. Is this dictated by color strings?   Do multi-parton interactions play a role?  Are there simply three or four or five clustered constituents to the nucleon?  Are there substantial contributions from the fluctuations in entropy deposition per constituent-constituent collision?   As shown in the PHENIX publication~\cite{new:phenix}, reproduced here as Table~\ref{tab:geometry}, the initial geometry of the collision systems has significant variation depending on the modeling of the aforementioned effects.   The template-, $c_1$- and ZYAM-adjusted values show a larger system-geometry difference in $v_2$ and a smaller system-geometry difference in $v_3$ compared to the raw results. Thus, the $v_2$ ($v_3$) adjusted values might indicate a smaller (larger) relative role for fluctuation-driven versus intrinsic geometry.

\begin{table}[hb]
\centering
\begin{ruledtabular}
\begin{tabular}{cccccc}
    Collision & Nucl. & Nucl. & Quarks & IP-G & IP-G \\
    System       & w/o  & w/ & w/  & w/  & w/  \\
     & NBD Fluc. &  NBD Fluc. &  NBD Fluc. &  Nucl. &  Quarks \\\hline
          & \multicolumn{5}{c}{$\left< \varepsilon_{2} \right>$} \\ \hline
            p+Au & 0.23 & 0.32 & 0.38 & 0.10 &  0.50 \\
           d+Au & 0.54 & 0.48 & 0.51 & 0.58 & 0.73 \\
           \heau & 0.50 & 0.50 & 0.52 & 0.55 & 0.64 \\ \hline
        & \multicolumn{5}{c}{$\left< \varepsilon_{3} \right>$} \\ \hline
           p+Au & 0.16 & 0.24 & 0.30 & 0.09 & 0.32 \\
           d+Au & 0.18 & 0.28 & 0.31 & 0.28 & 0.40 \\
           \heau & 0.28 & 0.32 &  0.35 &  0.34 & 0.46 \\
\end{tabular}
\end{ruledtabular}
\caption{Summary of various initial geometry calculations quantified by the average
  eccentricities $\varepsilon_{2(3)}$ in central (impact parameter $b<2~\text{fm}$) \pau,
  \dau, \heau events.  Column 2 uses Monte Carlo Glauber with nucleon position
  fluctuations~\cite{Nagle:2013lja}. Column 3 uses Monte Carlo Glauber with nucleon
  position fluctuations and negative binomial distribution (NBD) fluctuations in particle
  production~\cite{Welsh:2016siu}.  Column 4 uses Monte Carlo Glauber with constituent
  quark position fluctuations and NBD fluctuations~\cite{Welsh:2016siu}.  Columns 5 and 6
  use the \textsc{ip-glasma} framework with nucleon and constituent quark position fluctuations
  respectively, where both include gluon field fluctuations~\cite{Schenke:2012wb}.  These
  results were obtained with the publicly available \textsc{ip-glasma} code.
  \footnote{The eccentricities from \textsc{ip-glasma} depend on where the $Q_{s}^{2}$ Gaussian distribution in transverse coordinates is truncated.   For these values $r_\mathrm{max}$= 3~fm was utilized.}}
\label{tab:geometry}
\end{table}

\begin{figure}[!ht]
    \centering
        \includegraphics[width=0.9\linewidth]{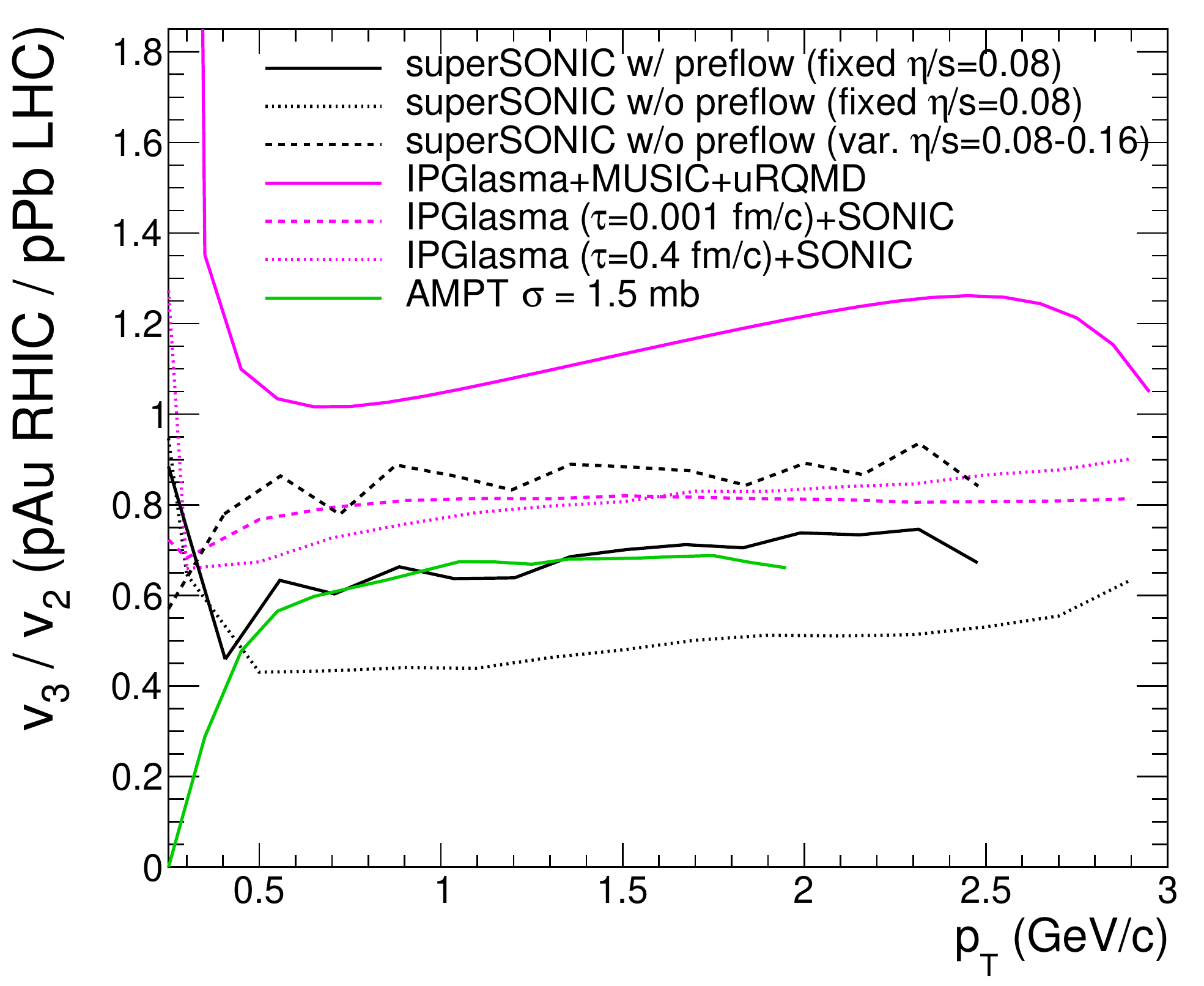}
    \caption{Various hydrodynamic and parton transport calculations with different initial states, pre-hydrodynamic modeling, and hydrodynamic evolution with results for the double ratio of $v_{3}/v_{2}$ in \pau collisions at \sqsn= 200 GeV over  $v_{3}/v_{2}$ in \ppb collisions at \sqsn= 5 TeV.  See text for details.}
    \label{fig:v2v3doubleratio}
\end{figure}

Another key effect is the physics of the pre-hydrodynamic stage (ii)---see the calculation differences between \textsc{sonic} without pre-flow  and \textsc{supersonic}  with strong pre-flow in Figure~\ref{fig:uber}.   

One can partially isolate this effect by comparing \pau collisions at RHIC and \ppb collisions at the LHC.  
If one makes the assumption that the initial geometry is very similar at both collision energies, the relative triangularity and ellipticity should be the same.    Figure~\ref{fig:v2v3doubleratio} shows various hydrodynamic and transport model calculations and the predicted double ratio of $v_{3}/v_{2}$ in \pau collisions at \sqsn= 200 GeV and \ppb collisions at \sqsn= 5 TeV.    All of the results for the double ratio are significantly below one, except for the \textsc{ip-glasma+music+urqmd} calculation~\cite{Schenke:2020mbo}, which is close to one.   The \textsc{ip-glasma} matching conditions to hydrodynamics or initial geometry between the collision energies may differ, though that needs to be confirmed.    

The reason for values near or significantly below one is that the smaller scale features of higher geometric moments ($\varepsilon_{n}$) take more time in the hydrodynamic stage to translate into flow moments $v_{n}$. Hence the lower-multiplicity, lower-initial-temperature hydrodynamic stage in collisions of lower energy at RHIC correlates with a shorter-lifetime hydrodynamic stage and thus a more striking
decrease in $v_3$ relative to $v_2$.    
The very low double ratio (0.4--0.5) in the \textsc{sonic} without pre-flow and fixed $\eta/s = 0.08$ and the much higher \textsc{supersonic} result (0.6--0.7) with pre-flow (modeled via AdS/CFT in the strongly coupled limit) is notable~\cite{Romatschke:2015gxa}. 
With a short hydrodynamic lifetime, particularly at RHIC, the additional push of the strongly-coupled pre-flow stage adds significantly to the translation of geometry to flow.   The calculation directly from Ref.~\cite{Romatschke:2015gxa} without pre-flow has an $\eta/s = 0.08$ at RHIC and $\eta/s = 0.16$ at the LHC---and this compensates to bring the double ratio up, though still significantly below one.   Calculations within the \textsc{ampt} framework also yield a result significantly below one.

\begin{figure}[!ht]
    \centering
    \includegraphics[width=0.9\linewidth]{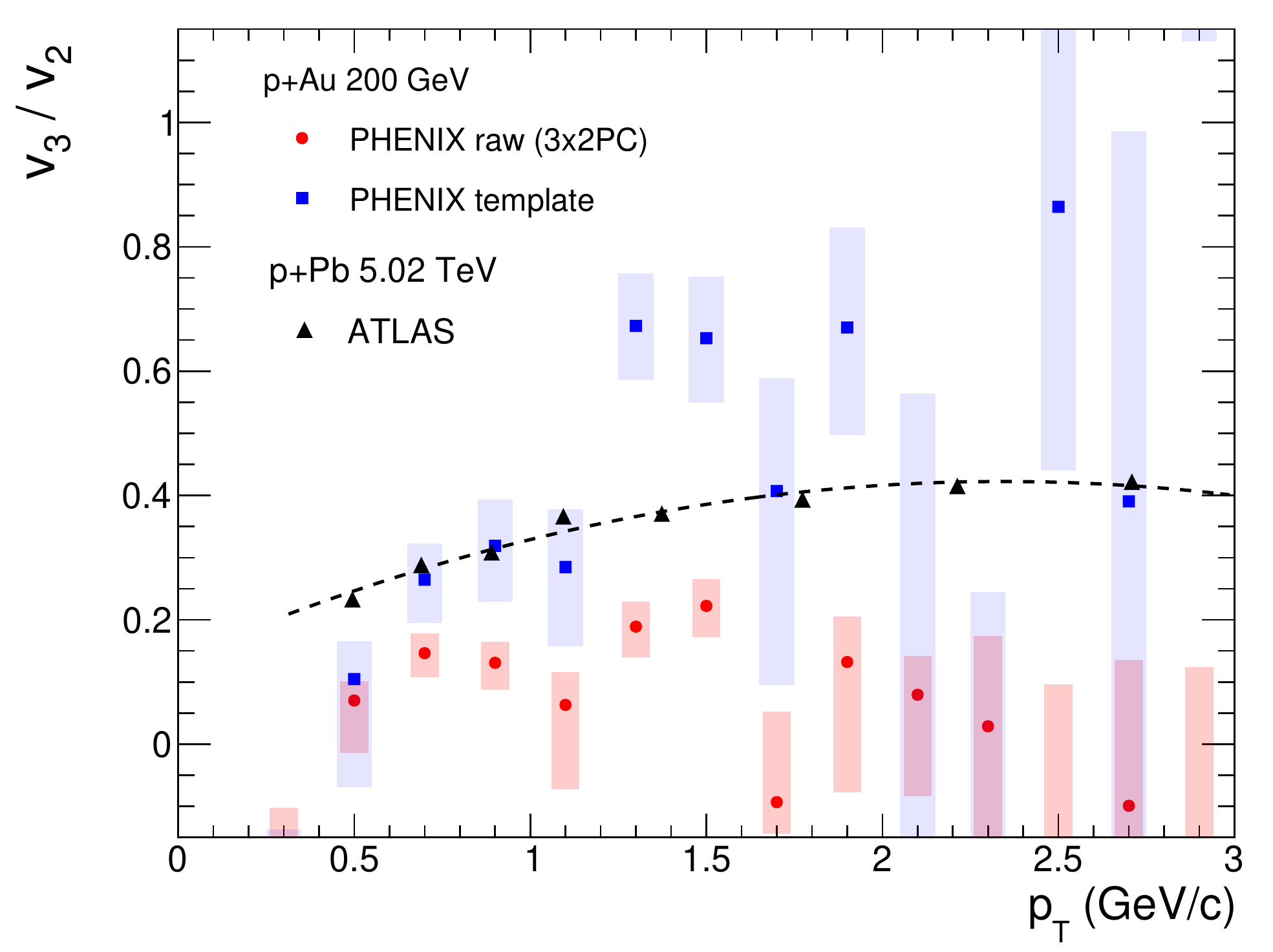}
    \includegraphics[width=0.9\linewidth]{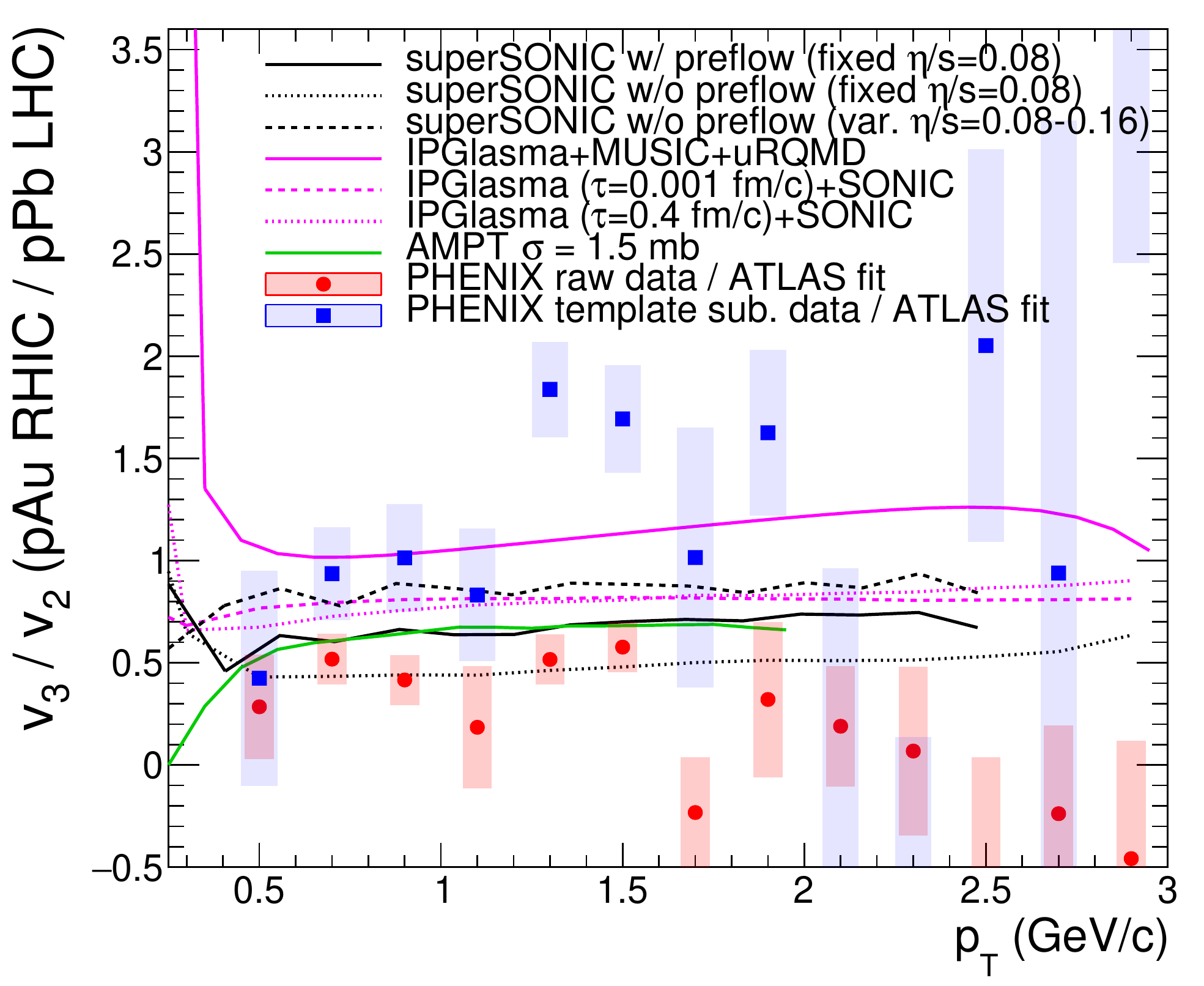}
    \caption{(Upper) Ratio of $v_3$/$v_2$ as a function of \pt from ATLAS \ppb collisions at \sqsn= 5 TeV and PHENIX raw and template adjusted \pau collisions at \sqsn= 200 GeV.   Uncertainties shown are statistical only.
    (Lower) Double ratio of PHENIX raw and template adjusted data over the ATLAS data fit compared with various theoretical calculations.}
    \label{fig:v3v2data}
\end{figure}

One can make the same type of comparison between LHC \ppb and RHIC \pau collisions with experimental data.   Figure~\ref{fig:v3v2data} (upper) shows the ATLAS published \ppb at \sqsn= 5 TeV data ratio for $v_3$/$v_2$ as a function of \pt.    The dashed line is a fit to the data.   The PHENIX \pau raw results from BBCS-FVTXS-CNT with statistical uncertainties only are shown, along with the template-adjusted values.       The raw \pau results are lower by a factor of 2--3 compared to the ATLAS \ppb values.    
Figure~\ref{fig:v3v2data} (lower) shows the ratio of PHENIX data to the ATLAS data fit compared with theoretical calculations.
The PHENIX raw data ratios are qualitatively consistent with the \textsc{sonic} calculations without pre-flow, which was utilized in the original \pdheau proposal paper~\cite{Nagle:2013lja}, though drop significantly at the highest \pt.    For the template-adjusted values for \pt $> 1.2$~GeV, the results are significantly higher than the ATLAS \ppb values (a factor of 1.5--2).    For \pt $<$~1.2~GeV, the PHENIX template-adjusted values are reasonably consistent with the ATLAS \ppb result, i.e., a double ratio near one.

It is highly probable, though not definitive, that the non-flow adjusted values are too low for $v_2$ and too high for $v_3$, i.e. an over-correction.
Without Monte Carlo with comparable flow and non-flow  contributions to real data, it is a challenge to further pin down the range of $v_3$ values reliably.   Thus, the issues of whether strong pre-flow in stage (ii) is needed and/or the intrinsic geometry relative to fluctuation-driven geometry is in a different balance for triangularity are still open.

\section{Collapse of Triangular Flow}

Within  the \textsc{(super)sonic} framework, triangular flow $v_3$ essentially collapses in \pa collisions below a particular collision energy~\cite{Romatschke:2015gxa}.    In the calculation, as the collision energy is reduced, what is really changing is the matching to the final hadron $dN_{ch}/d\eta$, which effectively shrinks the initial entropy and the length of time in the hydrodynamic phase, i.e. the lifetime of the QGP.  One can construct a map from $dN_{ch}/d\eta$ to $v_3$/$v_2$.    Since \textsc{supersonic} is 2+1D hydrodynamics, each collision energy is simply treated as a slice in pseudorapidity.   Thus, we have generated a map between collision energies and pseudorapidity in \pau collisions at \sqsn= 200 GeV.   

\begin{figure}[!ht]
    \centering
    \includegraphics[width=0.9\linewidth]{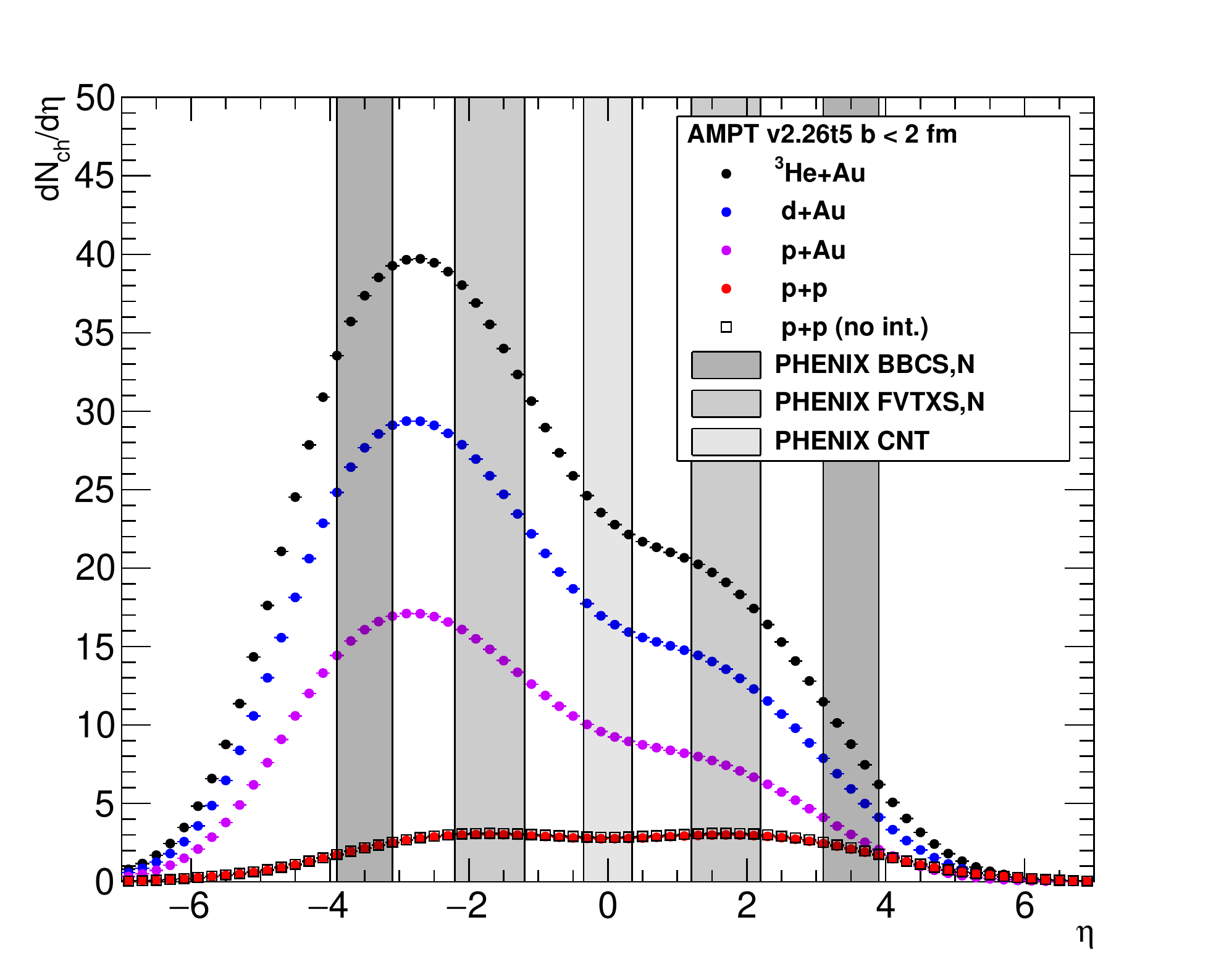}
    \caption{Results from \textsc{ampt} for $dN_{ch}/d\eta$ as a function of $\eta$ in minimum bias \pp (with and without final-state interactions) and central ($b < 2$~fm) \pau, \dau, and \heau collisions at \sqsn = 200 GeV.   Also highlighted are the approximate pseudorapidity acceptances for the PHENIX detector systems.}
    \label{fig:ampt_dndeta}
\end{figure}

Shown in Figure~\ref{fig:ampt_dndeta} is the pseudorapidity distribution $dN_{ch}/d\eta$ from \textsc{ampt} in minimum-bias \pp (with and without final-state interactions) and central ($b < 2$~fm) \pau, \dau, and \heau collisions at \sqsn~=~200~GeV.  The approximate pseudorapidity acceptances for the PHENIX detector systems are marked with boxes.   
Within the pseudorapidity range of the PHENIX measurement of $dN_{ch}/d\eta$ in these systems~\cite{Adare:2018toe}, \textsc{ampt} is in reasonable agreement with data.

\begin{figure}[!ht]
    \centering
    \includegraphics[width=0.9\linewidth]{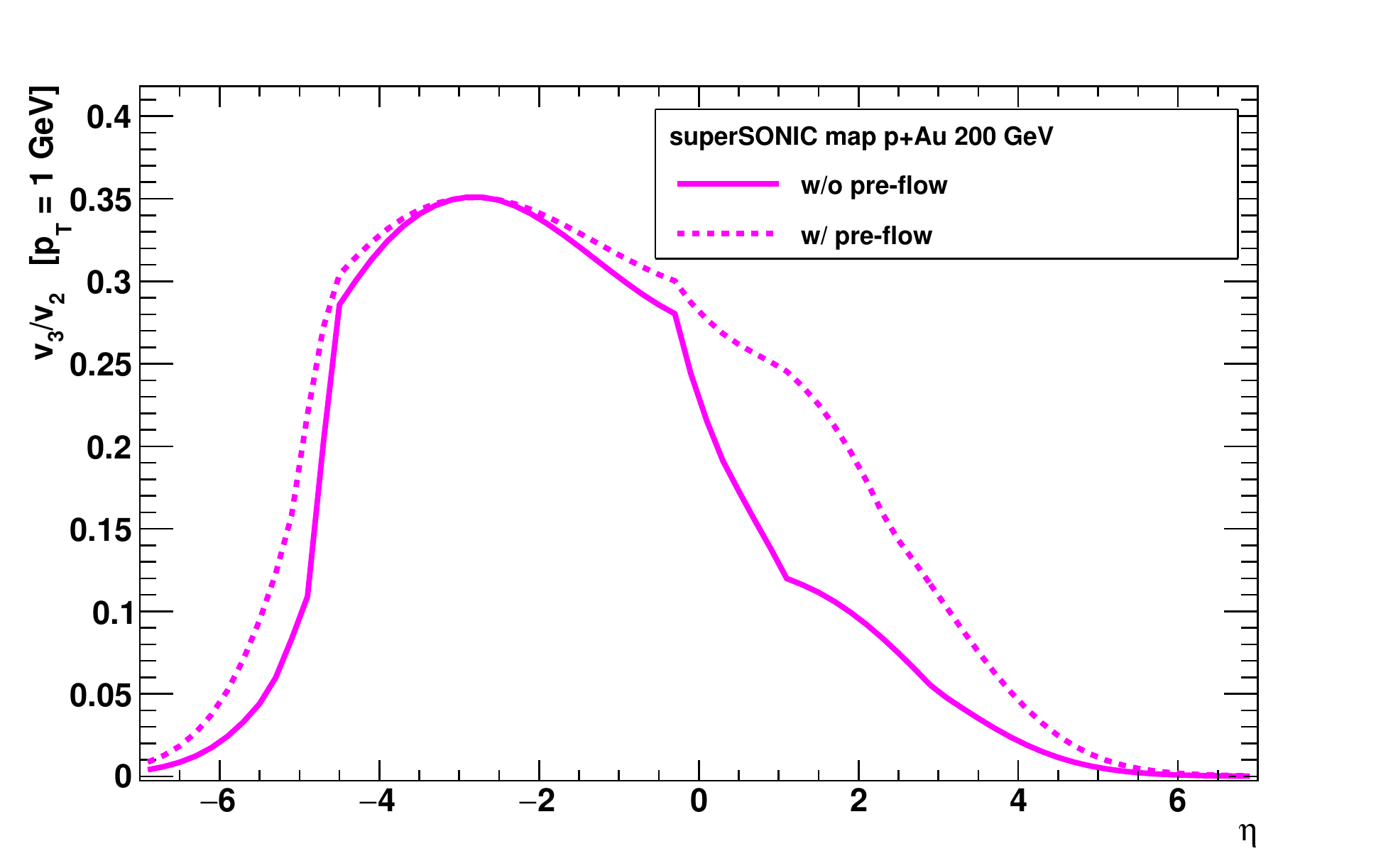}
    \caption{Theoretical predictions for the ratio $v_{3}/v_{2}$ at $p_{T}=1.0$~GeV as a function of pseudorapidity in \pau collisions at \sqsn = 200~GeV.    Results are shown for with (\textsc{supersonic}) and without (\textsc{sonic}) strongly-coupled pre-flow.    See text for calculation details.}
    \label{fig:supersonicmap}
\end{figure}

The \textsc{(super)sonic} results (with and without strongly coupled pre-flow) in \pa collisions as a function of \sqsn are given in Ref.~\cite{Romatschke:2015gxa}, including values of $dN_{ch}/d\eta$.   Using the \textsc{ampt} distribution of $dN_{ch}/d\eta(\eta)$, we have calculated the ratio $v_{3}/v_{2}$ as a function of pseudorapidity in \pau collisions at 200~GeV.   The resulting predictions are shown in Figure~\ref{fig:supersonicmap}.    There is a precipitous drop in the ratio when transitioning to forward pseudorapidity without pre-flow, which is expected from the same drop seen from \textsc{(super)sonic} in going from \pau at 200~GeV to 62.4~GeV---see Figure 4 from Ref.~\cite{Romatschke:2015gxa}.    The addition of strongly-coupled pre-flow mitigates this drop, though the ratio does decrease as the lifetime of the whole medium evolution shrinks in the forward rapidity slices.
These calculations put a spotlight on the important role of pre-hydrodynamic evolution and the importance in treating these asymmetric collisions systems asymmetrically, i.e. not assuming any symmetry even near mid-rapidity.    The calculations also highlight that non-flow effects, which are largest when $dN_{ch}/d\eta$ is smallest and when the real flow coefficient is smallest, could be highly pseudorapidity dependent.   A future experimental measurement of the $v_3/v_2$ ratio over a broad pseudorapidity range, though challenging, would be most instructive.

\section{Longitudinal Decorrelations}

So far we have focused on flow and non-flow contributions, but longitudinal decorrelations may also play a significant role.     If the entropy deposition in the transverse plane is dependent on the longitudinal or pseudorapidity slice, then the magnitude and orientation of the initial geometry, i.e. $\varepsilon_{n}$ and $\Psi_{n}$, respectively, may result in such flow decorrelations~\cite{PhysRevC.91.044904,Bozek:2017qir,Behera:2020mol}.    Longitudinal decorrelations have been measured in nucleus-nucleus collisions and these effects are larger for $v_3$ compared to $v_2$~\cite{Aad:2020gfz, Khachatryan:2015oea}.  The decorrelations effects are found to be larger at lower collision energies~\cite{Nie:2020trj,PhysRevC.97.024907}, and have not been quantified in small system collisions, notably \pdheau at RHIC.     Since the PHENIX results have detectors covering a range of pseudorapidities, the detector combinations used could be influenced by such decorrelations.

If we consider the parameterization of decorrelation used by the CMS Collaboration~\cite{Khachatryan:2015oea}, the correlation coefficients can be written as $c_{n} = v_{n,a} \times v_{n,b} \times \exp[-\alpha\Delta\eta]$.   In this case, the two-particle Fourier coefficient is now not simply the multiplicative product of the flow coefficient for particle $a$ and particle $b$, but includes a longitudinal decorrelation proportional to the exponential of the pseudorapidity gap between particles $a$ and $b$ and a proportionality constant $\alpha$.    While it is unclear if this parameterization holds in \pdheau collisions over a wide range in pseudorapidity, it is nonetheless useful to explore the implications.   In the case of the BBCS-FVTXS-CNT detector combination, the $\Delta \eta$ values are approximately 3.5, 1.75, 1.75 for the BBCS-CNT, BBCS-FVTXS, FVTXS-CNT, respectively.   One can express this as below
\begin{widetext}
\begin{align}
v_n(\text{CNT}) &= \sqrt{{c_n[\text{BBCS-CNT}] \times c_n[\text{FVTXS-CNT}]}\over{c_n[\text{BBCS-FVTXS}]}} \\
                &= \sqrt{{v_n[\text{BBC}] v_n[\text{CNT}] e^{-3.5\alpha} \times v_n[\text{FVTXS}] v_n[\text{CNT}]  e^{-1.75\alpha}} \over{v_n[\text{BBCS}] v_n[\text{FVTXS}] e^{-1.75\alpha}}} \\
                &= v_n[\text{CNT}] e^{-3.5\alpha/2}
\end{align}
\end{widetext}
and one finds that in this case the decorrelation between the BBCS-FVTXS and FVTXS-CNT cancels, and only the square-root of the BBCS-CNT decorrelation remains.     Thus, $\alpha = 0.054$ would correspond to a 10\% decorrelation in $c_n$ over a two unit rapidity gap, corresponding to a 9\% lower $v_{n}$ being measured by the BBCS-FVTXS-CNT combination.   In contrast, in the case of FVTXS-CNT-FVTXN, all of the decorrelation cancels out in that combination.    Of course, this all assumes this simple exponential model.   More detailed modeling of longitudinal decorrelation effects in small systems might lend more insights.

\section{Summary}

Utilizing the PHENIX published correlation coefficients, we have tested various flow and non-flow adjustment methods.    The results vary depending on the method quantitatively and we have discussed potential implications in light of significant non-closure results with \textsc{ampt} and \textsc{pythia/angantyr}.  Comparisons between \pau at RHIC and \ppb at the LHC elucidate the potential influence of pre-hydrodynamic evolution, via comparisons with \textsc{(super)sonic} calculations.
The conclusion that these flow coefficients are dominated by initial geometry coupled with final-state interactions (e.g.~hydrodynamic expansion of quark-gluon plasma) is confirmed, and explanations based on initial-state glasma are ruled out.     The detailed balance of intrinsic geometry and fluctuation-driven geometry as well as the exact role of weak or strong coupled pre-hydrodynamic evolution remain open questions requiring further theoretical and experimental investigation.  

\section{Acknowledgments}

We thank the PHENIX Collaboration for useful, open discussions and suggestions leading to this manuscript.
We thank the Brookhaven National Laboratory Small System Task Force (Constantin Loizides, Jean-Yves Ollitrault, Sergei Voloshin) for useful input.  We acknowledge useful discussions with Paul Romatschke, Julia Velkovska, and Bill Zajc.  

JLN acknowledges support from the U.S. Department of Energy, Office of Science, Office of Nuclear Physics under Contract No. DE-FG02-00ER41152.
SHL acknowledges support from the National Research Foundation of Korea (NRF) grant
funded by the Korea government (MSIT) under Contract No. 2020R1C1C1004985.

\newpage

\bibliography{main}

\end{document}